\documentclass[a4paper,11pt]{article}
\pdfoutput=1 

\usepackage{jheppub} 

\usepackage[T1]{fontenc} 
\usepackage{lmodern}
\usepackage{cancel}
\usepackage{arydshln}
\usepackage{dsfont}
\usepackage{subcaption}
\usepackage{caption}
\usepackage{xcolor}
\newcommand{\be}{\begin{equation}}
\newcommand{\ee}{\end{equation}}

\usepackage{appendix}
\usepackage{orcidlink}
\usepackage{tikz-feynman}
\usepackage[inkscapelatex=false]{svg}
\DeclareUnicodeCharacter{2212}{-}

\title{\boldmath Tensor Meson Pole contributions to the HLbL piece of $a_{\mu}^{\rm{HLbL}}$ within $R \chi T$}

\author[a]{Emilio J. Estrada,}
\author[a,b]{and Pablo Roig}
\affiliation[a]{Departamento de F\'isica, Centro de Investigaci\'{o}n y de Estudios Avanzados del Instituto Polit\'{e}cnico Nacional, Apdo. Postal 14-740, 07000 Ciudad de M\'{e}xico, M\'{e}xico}
\affiliation[b]{IFIC, Universitat de Val\`{e}ncia – CSIC, Catedr\'atico Jos\'e Beltr\'an 2, E-46980 Paterna, Spain}

\emailAdd{emilio.estrada@cinvestav.mx}
\emailAdd{pablo.roig@cinvestav.mx}

\abstract{We compute the tensor meson pole contributions to the Hadronic Light-by-Light piece of $a_\mu$ in the purely hadronic region, using Resonance Chiral Theory. Given the differences between the dispersive and holographic 
groups determinations and the resulting discussion of the corresponding uncertainty estimate for the Hadronic Light-by-Light section of the muon \ensuremath{g - 2} theory initiative second White Paper, we consider timely to present an alternative evaluation. In our approach, in addition to the lightest tensor meson nonet, two vector meson resonance nonets are considered, in the chiral limit.Disregarding operators with derivatives, only the form factor $\mathcal{F}_1^T$ is non-vanishing, as assumed in the dispersive study. All parameters are determined by imposing a set of short-distance QCD constraints, and the radiative tensor decay widths. In this case, we obtain the following results for the different contributions (in units of $10^{-11}$): $a_\mu^{\rm a_2-pole}=-\left(1.02(10)_{\rm stat}(^{+0.00}_{-0.12})_{\rm syst}\right)$, $a_\mu^{\rm f_2-pole}=-\left(3.2(3)_{\rm stat}(^{+0.0}_{-0.4})_{\rm syst}\right)$ and $a_\mu^{\rm f_2^\prime-pole}=-\left(0.042(13)_{\rm stat}\right)\,$
, which add up to $a_\mu^{a_2+f_2+f_2^\prime \rm - pole}=-\left(4.3^{+0.3}_{-0.5}\right)$, in close agreement with the holographic result when truncated to $\mathcal{F}_1^T$ only. However, with an ad-hoc extended Lagrangian, that also generates $\mathcal{F}_3^T$, as in the holographic approach, we have found: $a_\mu^{\rm a_2-pole}=+0.47(1.43)_{\rm norm}(3)_{\rm stat}(^{+0.06}_{-0.00})_{\rm syst}$, $a_\mu^{\rm f_2-pole}=+1.18(4.18)_{\rm norm}(12)_{\rm stat}(^{+0.24}_{-0.00})_{\rm syst}$ and $a_\mu^{\rm f_2'-pole}=$ $+0.040(78)_{\rm norm}(2)_{\rm stat}$, summing to $a_\mu^{a_2+f_2+f_2^\prime \rm - pole}=+1.7(4.4)$, which agree 
with these recent determinations within uncertainties (dominated by the $\mathcal{F}_3^T$ normalization). We point out that $R\chi T$ generates all five form factors, differently to previous approaches. The contributions to $a_\mu$ of $\mathcal{F}_{2,4,5}$ cannot be evaluated in the current basis, preventing for the moment a complete calculation of $a_\mu^{\rm T-poles}$ within our framework.}

\begin{document} 
\maketitle
\flushbottom

\section{Introduction}
The final measurement of the Muon \ensuremath{g - 2} Collaboration will be announced soon and its prediction in the Standard Model will be updated just before~\footnote{It appeared during the referral of this paper \protect\cite{Aliberti:2025beg}, with an increased uncertainty of $62\times10^{-11}$.}, marking a milestone in the exploration of the high-intensity frontier of particle physics.

The current experimental average is based on the legacy BNL result \cite{Muong-2:2006rrc} and in the two measurements published so far by FNAL \cite{Muong-2:2021ojo,Muong-2:2023cdq}, reaching an uncertainty of $22\times10^{-11}$~\cite{ParticleDataGroup:2024cfk}~\footnote{It was reduced to $15\times10^{-11}$ thanks to the final FNAL results~\cite{Muong-2:2025xyk}, which were made public during the referral of this work.}. The current SM prediction was published in the White Paper \cite{Aoyama:2020ynm,Colangelo:2022jxc}, based on refs.~\cite{Davier:2017zfy,Keshavarzi:2018mgv,Colangelo:2018mtw,Hoferichter:2019mqg,Davier:2019can,Keshavarzi:2019abf,Kurz:2014wya,FermilabLattice:2017wgj,Budapest-Marseille-Wuppertal:2017okr,RBC:2018dos,Giusti:2019xct,Shintani:2019wai,FermilabLattice:2019ugu,Gerardin:2019rua,Aubin:2019usy,Giusti:2019hkz,Melnikov:2003xd,Masjuan:2017tvw,Colangelo:2017fiz,Hoferichter:2018kwz,Gerardin:2019vio,Bijnens:2019ghy,Colangelo:2019uex,Pauk:2014rta,Danilkin:2016hnh,Jegerlehner:2017gek,Knecht:2018sci,Eichmann:2019bqf,Roig:2019reh,Colangelo:2014qya,Blum:2019ugy,Aoyama:2012wk,Aoyama:2019ryr,Czarnecki:2002nt,Gnendiger:2013pva}~\footnote{See also, e. g. refs.~\cite{Borsanyi:2020mff,Lehner:2020crt,Knecht:2020xyr,Masjuan:2020jsf,Ludtke:2020moa,Chao:2020kwq,Miranda:2020wdg,Hoid:2020xjs,Bijnens:2020xnl,Colangelo:2020lcg,Bijnens:2021jqo,Bijnens:2022itw,Bijnens:2024jgh,Chao:2021tvp,Danilkin:2021icn,Colangelo:2021nkr,Leutgeb:2021mpu,Colangelo:2021moe,Cappiello:2021vzi,Giusti:2021dvd,Hoferichter:2021wyj,Miramontes:2021exi,Raya:2019dnh,Stamen:2022uqh,Boito:2022rkw,Wang:2022lkq,Aubin:2022hgm,Colangelo:2022vok,Ce:2022kxy,ExtendedTwistedMass:2022jpw,Colangelo:2022prz,Biloshytskyi:2022ets,Boito:2022dry,Leutgeb:2022lqw,CMD-3:2023alj,CMD-3:2023rfe,Hoferichter:2023bjm,Colangelo:2023een,FermilabLatticeHPQCD:2023jof,RBC:2023pvn,Wang:2023njt,Ludtke:2023hvz,Escribano:2023seb,Blum:2023vlm,Masjuan:2023qsp,Benton:2023dci,Hoferichter:2023sli,Davier:2023cyp,Davier:2023fpl,Hoferichter:2024fsj,Boccaletti:2024guq,Lahert:2024vvu,Castro:2024prg,Hoferichter:2025yih,FermilabLatticeHPQCD:2024ppc,Holz:2024diw,Leutgeb:2024rfs,Holz:2024lom,Miramontes:2024fgo,Deineka:2024mzt,Estrada:2024rfi,Volkov:2019phy,Volkov:2024yzc,Aoyama:2024aly}.}, with an error of $43\times10^{-11}$. A reduction of this uncertainty at the level of $\leq16\times10^{-11}$ would be desirable to maximize the new physics implications of the final FNAL result~\cite{Muong-2:2015xgu}.
Within this global effort, we will focus in this paper on a subleading contribution to the hadronic light-by-light (HLbL) piece of the muon anomaly ($a_\mu=(g_\mu-2)/2$) which seems puzzling at present.

A tension has recently appeared, between the determinations of the tensor meson contributions to the HLbL piece of $a_\mu$ by the dispersive \cite{Hoferichter:2024bae,Hoferichter:2024vbu} and holographic\cite{Cappiello:2025fyf,Mager:2025pvz} groups (see also the earlier holographic computation of ref.~\cite{Colangelo:2024xfh} and refs.~\cite{Pascalutsa:2012pr,Pauk:2014rta, Danilkin:2016hnh,Jegerlehner:2017gek}, which were used to estimate it in the first White Paper~\cite{Aoyama:2020ynm}). While these results basically agree on its magnitude, they differ in sign. Interestingly, the holographic result would bring into close agreement the phenomenological determination with the lattice result for the whole $a_\mu^{\rm HLbL}$. Moreover, the contribution of the tensor meson poles in the dispersive study relies on a quark model FF, which may need to be refined. This current puzzle motivated us to compute the dispersively defined lightest tensor meson pole contributions within Resonance Chiral Theory ($R\chi T$, see section \ref{sec:2})~\cite{Ecker:1988te,Ecker:1989yg}, which bridges between the chiral and asymptotic QCD regimes.

The study of tensor mesons in $R\chi T$ is scarce, as it contributes negligibly to chiral dynamics \cite{Ecker:2007us} (see also ref.~\cite{Bellucci:1994eb}). We will benefit from the recent ref.~\cite{Chen:2023ybr}, which particularly focused on understanding tensor meson masses and their di-pion and radiative decay widths.

Neglecting --by assumption-- operators with derivatives, we have found that only $\mathcal{F}_1^T$ is nonvanishing, as in the dispersive group analysis \cite{Hoferichter:2024bae}. This is important, since one of such FFs, $\mathcal{F}_3^T$, appears responsible for the positive sign found in the holographic computation~\cite{Cappiello:2025fyf} of this contribution to $a_\mu$. Our description of the $T\gamma\gamma$ interactions fulfills the short-distance QCD constraint on the tensor mesons transition form factors (TFFs) $\mathcal{F}_1^T$ when both virtualities are asymptotic and in the Brodsky-Lepage limit~\cite{Brodsky:1973kr,Hoferichter:2020lap}. This requires a $\sim 1/Q^4$ damping of the $\mathcal{F}_1^T$ FF in the ultraviolet, which is achieved if at least two vector meson multiplets of resonances are accounted for. If we enlarge our Lagrangian, allowing for a selected set of operators with two derivatives that generate contributions to $\mathcal{F}_3^T$ only, we get short-distance constraints independent of those from $\mathcal{F}_1^T$. In this case, we reproduce the sign found in the holographic calculation 
we found results compatible within errors with both approaches
, with our large uncertainty dominated by the normalization of $\mathcal{F}_3^T$. We point out that non-vanishing $\mathcal F_{1,\dots,5}^T$ are naturally generated within $R\chi T$. Computing their contribution to $a_\mu$ (as required in a complete $R\chi T$ evaluation of $a_\mu^{\rm T-poles}$) would need a new basis, which we put forward as a necessary development --beyond our particular framework-- of the existing formalism.

The outline of this paper follows. We describe our effective approach in section \ref{sec:2}, impose high-energy constraints on the obtained FFs in section \ref{sec:3} and finally present our results --discussing the uncertainty estimation-- for the $a_\mu^{\rm HLbL:\, T-poles}$ in section \ref{sec:4}. We conclude in section \ref{sec:Concl}. An \nameref{sec:appendix} explains how we estimated the normalization of $\mathcal{F}_3$, which saturates the uncertainty of our results.

\section{Resonance Chiral Theory Lagrangian}
Our theoretical framework will be a Lagrangian theory, that is briefly introduced in the following. In order to extend the applicability of Chiral Perturbation Theory \cite{Weinberg:1978kz, Gasser:1983yg, Gasser:1984gg} to larger energies, Resonance Chiral Theory ($R\chi T$) \cite{Ecker:1988te, Ecker:1989yg} includes the resonances as dynamical fields. Independence on the formalism used to describe the spin-one mesons is verified for the two-point functions and related observables, and the convenience of using antisymmetric tensors for the spin-one fields is explained~\cite{Ecker:1988te, Ecker:1989yg}. $R\chi T$ is built upon approximate flavor symmetries: chiral for the pseudo-Goldstone bosons (pion, kaon and eta mesons) and unitary for the resonances. Its expansion parameter is the inverse of the number of colors of the QCD gauge group~\cite{tHooft:1973alw}. The number of relevant operators is limited by two observations. On the one hand, in $R\chi T$ the number of resonance fields is typically restricted by the specific problem at hand. On the other, the inclusion of tensors of high order in the chiral counting may conflict with the requirements from short-distance QCD. Particularly, for order parameters of chiral symmetry breaking, the matching of $R\chi T$ results to the corresponding operator product expansion (OPE) expressions has been pursued \cite{Ecker:1988te, Ecker:1989yg,Cirigliano:2004ue,Cirigliano:2006hb,Cirigliano:2005xn,Dai:2019lmj,Ruiz-Femenia:2003jdx,Kampf:2011ty,Roig:2013baa}, rendering relations between Lagrangian parameters, thus increasing the predictive power of this framework. $R\chi T$ has been employed successfully in the computation of a number of hadronic contributions to $a_\mu$~\cite{Kampf:2011ty,Roig:2014uja,Guevara:2018rhj,Kadavy:2022scu, Estrada:2024cfy,Roig:2019reh,Pich:2001pj, Cirigliano:2001er, Cirigliano:2002pv,Miranda:2020wdg,Qin:2020udp,Masjuan:2023qsp,Castro:2024prg,Wang:2023njt,Qin:2024ulb}.

We have considered the most general Lagrangian consistent with the symmetries of $R \chi T$, at leading order in the $1/N_C$ expansion, in the chiral limit ($m_{u,d,s}\to 0$) and --by assumption-- without derivatives (we reconsidered this hypothesis in Table \ref{tab:operatorsNLO}). In the minimal realization, which includes pseudoscalar mesons, vector meson resonances, photons, tensor mesons, and their interactions, only 6 operators contribute -- in this approximation-- to the radiative tensor decays, which are collected in Table \ref{tab:operators}. We further considered an ad hoc extension consisting in the operators of the kind $\langle T^{\mu\nu} \{\nabla_\alpha R^{\alpha}_\mu,\nabla_\lambda R^{(\prime) \lambda}_\nu\}\rangle$~\footnote{We choose this particular extension since it generates a non-zero $F_T^3$ TFF, as put forward in \cite{Cappiello:2025fyf}. This allows for an evaluation of $a_\mu^{\rm HLbL: \, T-poles}$ which is free of kinematic singularities in the optimized basis of Ref. \cite{Hoferichter:2024fsj} (contrary to what happens in a more general treatment, that deserves further study).} --with $R,R^\prime=f,V,V^\prime$ as in Table \ref{tab:operators}--, resulting in 6 additional operators, which are displayed in Table \ref{tab:operatorsNLO}~\footnote{The covariant derivatives are constructed as in refs. \cite{Gasser:1984gg,Ecker:1988te}. However, for the case at hand, $\nabla_\lambda \to \partial_\lambda$.}. 
\begin{table}[t!]
    \centering
    \begin{tabular}{|c| c |}\hline
    Coupling constant & Operator \\ \hline
    $C_{T\gamma\gamma}$   & $\langle T^{\mu\nu} \{f_{+\mu}^{\alpha},f_{+\alpha \nu}\}\rangle$ \\
    $C_{T\gamma V}$   & $i\langle T^{\mu\nu}\{ f_{+\mu}^{\alpha},V_{\alpha\nu}\}\rangle$ \\
    $C_{T\gamma V^\prime}$   & $i\langle T^{\mu\nu}\{ f_{+\mu}^{\alpha},V^{\prime}_{\alpha\nu}\}\rangle$ \\
    $C_{TVV}$   & $\langle T^{\mu\nu} \{V_{\mu}^{\alpha},V_{\alpha \nu}\}\rangle$ \\
    $C_{TV^\prime V^\prime}$   & $\langle T^{\mu\nu} \{V_{\mu}^{\prime\alpha},V^{\prime}_{\alpha \nu}\}\rangle$ \\
    $C_{TV V^\prime}$   & $\langle T^{\mu\nu} \{V_{\mu}^{\alpha},V^{\prime}_{\alpha \nu}\}\rangle$ \\\hline
    \end{tabular}
    \caption{Relevant $R \chi T$ Operators at leading order in the $1/N_C$ expansion and in the chiral limit, neglecting operators with derivatives. We note that $C_{T\gamma V^{(\prime)}}$ are dimensionless, while $(C_{T\gamma\gamma})$ $C_{TV^{(\prime)}V^{(\prime)}}$ have (inverse) energy dimensions.}
    \label{tab:operators}
\end{table}
\begin{table}[t!]
    \centering
    \begin{tabular}{|c| c |}\hline
    Coupling constant & Operator \\ \hline
$g_{T\gamma\gamma}$   & $\langle T^{\mu\nu} \{\nabla_{\alpha}f_{+\mu}^{\alpha},\nabla_\lambda f^{\lambda}_{+ \nu}\}\rangle$\\
$g_{T\gamma V}$   & $i \langle T^{\mu\nu} \{\nabla_{\alpha}V_{\mu}^{\alpha},\nabla_\lambda f^{\lambda}_{+ \nu}\}\rangle$ \\
$g_{T\gamma V^\prime}$   & $i\langle T^{\mu\nu} \{\nabla_{\alpha}V_{\mu}^{\prime\,\alpha},\nabla_\lambda f^{\lambda}_{+ \nu}\}\rangle$ \\
$g_{TVV}$   & $\langle T^{\mu\nu} \{\nabla_{\alpha}V_{\mu}^{\alpha},\nabla_\lambda V^{\lambda}_{ \nu}\}\rangle$ \\
$g_{TV^\prime V^\prime}$   & $\langle T^{\mu\nu} \{\nabla_{\alpha}V_{\mu}^{\prime\,\alpha},\nabla_\lambda V^{\prime\,\lambda}_{ \nu}\}\rangle$ \\
$g_{TV V^\prime}$   & $\langle T^{\mu\nu} \{\nabla_{\alpha}V_{\mu}^{\alpha},\nabla_\lambda V^{\prime\,\lambda}_{ \nu}\}\rangle$ \\\hline
    \end{tabular}
    \caption{$R \chi T$ Operators that are obtained adding two derivatives to those in Table~\ref{tab:operators}, and only generate $\mathcal{F}_T^3$. The energy dimension of the coefficients of the operators with zero, one and two resonances is -3, -2 and -1, respectively.}
    \label{tab:operatorsNLO}
\end{table}
Their real coupling constants are named according to the interaction vertices they describe and are not fixed by symmetry requirements.

We recall the $V\to\gamma$ terms (with trivial generalization for $V\to V^{(')}$), given in the original $R\chi T$ Lagrangian~\cite{Ecker:1988te}, that are also needed to describe the tensor meson transition FFs,
\begin{equation}
    \mathcal{L}^{V\to \gamma}=\frac{F_V}{2\sqrt{2}}\langle V_{\mu\nu}f_+^{\mu\nu} \rangle.
\end{equation}
The tensor mesons, $T^{\mu\nu}$, in the symmetric, on-shell traceless representation,  are~\cite{Ecker:2007us}:
\begin{equation}
    T_{\mu\nu}
    =\begin{pmatrix}
        \frac{a_2^0}{\sqrt{2}}+\frac{f^8_2}{\sqrt{6}}+\frac{f^0_2}{\sqrt{3}} & a_2^+& K_2^{*+}\\
        a_2^-& -\frac{a_2^0}{\sqrt{2}}+\frac{f^8_2}{\sqrt{6}}+\frac{f^0_2}{\sqrt{3}} & K_2^{*0}\\
        K_2^{*-}& \bar{K}_2^{*0}& -\frac{2 f_2^{8}}{\sqrt{6}}+\frac{f_2^{0}}{\sqrt{3}}
    \end{pmatrix}_{\mu\nu},
\end{equation}
where the $f_2^8$ and $f_2^0$ octet and singlet states in flavor space, respectively, are related to the physical mass eigenstates via a single-angle mixing:
\begin{gather}
    f_2^{8}=\sin\theta_T f_2+\cos\theta_T f_2^\prime,\nonumber \\
    f_2^{0}=\cos\theta_T f_2 - \sin\theta_T f_2^\prime.
\end{gather}

The vector meson nonets are in the antisymmetric tensor representation~\cite{Ecker:1988te}. Considering ideal mixing, they read
\begin{equation}
    V_{\mu\nu}
    =\begin{pmatrix}
        (\rho^0+\omega)/\sqrt{2} & \rho^+&K^{*+}\\
        \rho^{-}& (-\rho^0+\omega)/\sqrt{2} & K^{*0}\\
        K^{*-}&\bar{K}^{*0}&\phi
    \end{pmatrix}_{\mu\nu},
\end{equation}

and similarly for $V\to V^\prime$. The photon field is encoded in the $f_+$ chiral tensor~\cite{Bijnens:1999sh}:
\begin{equation}
    f_+^{\mu\nu}=uF_{L}^{\mu\nu}u^{\dagger}+u^\dagger F_R^{\mu\nu}u,\qquad F_{L(R)}^{\mu\nu}=-eQF^{\mu\nu}+...,
\end{equation}
where $Q=\rm{diag}(2/3,-1/3,-1/3)$, and $u=\exp\left(i\frac{\Phi}{\sqrt{2}F}\right)$ holds the information of the pseudo-Goldstone bosons ($\pi$, kaon and $\eta$ mesons), which in this case do not play a role. Given these building blocks, we can construct the amplitude of the $T\to \gamma \gamma$ decays and examine the associated transition FF (see Figure \ref{fig:TTFF}).
\begin{figure}[t!]
    \centering      
        \begin{tikzpicture}
        \begin{feynman}
        \vertex (a){\(T^{\alpha\beta}\)};
        \vertex[blob, right=of a](b){};
        \vertex[above right=of b](c1){\(\gamma*\)};
        \vertex[below right=of b](c2){\(\gamma*\)};        
        \diagram* {
            (a) --  (b)-- [photon] (c1),
            (b)-- [photon] (c2),
        };
        \end{feynman}
        \end{tikzpicture}
    \caption{Tensor Meson transition FF, which gives the radiative decay with both photons on-shell.}
    \label{fig:TTFF}
\end{figure}

We recall the matrix element of a massive tensor meson decaying into two virtual photons~\cite{Hoferichter:2024bae}
\begin{equation}\label{eq_me}
\langle\gamma^*(q_1,\lambda_1)\gamma^*(q_2,\lambda_2)|T(p,\lambda_T)\rangle=i(2\pi)^4\delta^{(4)}(q_1+q_2-p)e^2\varepsilon_{\mu}^{\lambda_1^*}(q_1)\varepsilon_{\nu} ^{\lambda_2^*}(q_2)\varepsilon^{\lambda_T}_{\alpha\beta}(p)\mathcal{M}^{\mu\nu\alpha\beta}(q_1,\,q_2)\,.
\end{equation}

For the tensor mesons, the reduced amplitude (which -among other properties~\cite{Hoferichter:2024bae}- must be symmetric under $q_1\leftrightarrow q_2,\mu\leftrightarrow\nu$) can be split into 5 structures contributing to observables~\cite{Hoferichter:2020lap}:
\begin{equation}
    \mathcal{M}^{\mu\nu\alpha\beta}=\sum_{i=5}^{5}T_i^{\mu\nu\alpha\beta}\frac{1}{m_T^{n_i}}\mathcal{F}_i^T(q_1^2,q_2^2),
\end{equation}
where $n_1=1$ and $n_{i\neq1}=3$, and the tensor structures are:
\begin{gather}
    T_1^{\mu\nu\alpha\beta}=g^{\mu\alpha }P_{21}^{\nu\beta}+g^{\nu\alpha} P_{12}^{\mu\beta}+g^{\mu\beta} P_{21}^{\nu\alpha}+g^{\nu\beta} P_{12}^{\mu\alpha}+g^{\mu\nu} (q_1^\alpha q_2^\beta+q_1^\beta q_2^\alpha)-q_1 \cdot q_2 (g^{\mu\alpha} g^{\nu\beta}+g^{\nu\alpha}g^{\mu\beta}),\nonumber \\
    T_2^{\mu\nu\alpha\beta}=(q_1^\alpha q_1^\beta+q_2^\alpha q_2^\beta)P_{12}^{\mu\nu}, \nonumber\\
    T_3^{\mu\nu\alpha\beta}=P_{11}^{\mu\alpha}P_{22}^{\nu\beta}+P_{11}^{\mu\beta}P_{22}^{\nu\alpha},\\
    T_4^{\mu\nu\alpha\beta}=P_{12}^{\mu\alpha}P_{22}^{\nu\beta}+P_{12}^{\mu\beta}P_{22}^{\nu\alpha},\nonumber\\
    T_{5}^{\mu\nu\alpha\beta}=P_{21}^{\nu\alpha}P_{11}^{\mu\beta}+P_{21}^{\nu\beta}P_{11}^{\mu\alpha},\nonumber
\end{gather}
where the $P_{ij}^{\mu\nu}$ are defined as:
\begin{equation}
    P_{ij}^{\mu\nu}=g^{\mu\nu}q_i \cdot q_j - q_i^\nu q_j^{\mu}.
\end{equation}

Using just the tensor meson interactions from Table \ref{tab:operators},
 only $\mathcal{F}_1^T$ appears in the $T\to \gamma^*\gamma ^*$ amplitude, as in ref. \cite{Hoferichter:2024bae}. However, when including the operators in Table \ref{tab:operatorsNLO}, we also get a non-vanishing $\mathcal{F}_3^T$~\footnote{ A study, within $R \chi T$, of the 5 TFFs generated with all the operators adding two derivatives to those in Table \ref{tab:operators} is possible. This is a work in preparation by the authors\cite{EmilioyPabloWIP}.}, in agreement with the holographic outcome \cite{Cappiello:2025fyf}
. The relative sign between the normalization of both FFs is fixed from the chiral expansion of the results in ref.~\cite{Cappiello:2025fyf} and determines the positive sign of the $a_\mu^{\rm T-pole}$ contribution, in agreement with the holographic result~\cite{Cappiello:2025fyf}.
\label{sec:2}
\section{$\mathcal{F}_{1,3}^{T}$ and short-distance constraints}
\label{sec:3}
From the operators in Table \ref{tab:operators}, we get the following transition FF for the 3 contributing neutral tensor mesons:
\begin{equation}
\begin{split}
\mathcal{F}_1^{a_2}(q_1^2,q_2^2)=&-\frac{M_{a_2}}{2}\Bigg[\frac{
\sqrt{2}}{3}C_{T\gamma\gamma}+C_{T\gamma V}
\frac{4 F_V}{3}\left(D^{-1}_{M_V}(q_1^2)+D^{-1}_{M_V}(q_2^2)\right)
\\
& 
+C_{TVV}
\frac{2\sqrt{2} F_V^2}{3} D^{-1}_{M_V}(q_1^2)D^{-1}_{M_V}(q_2^2)
+C_{TVV^\prime}
\frac{2\sqrt{2}F_V F_{V^\prime}}{3}
D^{-1}_{M_V}(q_1^2)D^{-1}_{M_{V^\prime}}(q_2^2)
+V\leftrightarrow V^\prime\Bigg],
\end{split}
\end{equation}
\begin{equation}
\begin{split}
\mathcal{F}_1^{f_2}(q_1^2,q_2^2)=&-\frac{M_{f_2}}{2}\Bigg[\frac{
\sqrt{2}}{3\sqrt{3}}C_{T\gamma\gamma}\left(\sin \theta_T + 2\sqrt{2}\cos \theta_T\right)\\
&+C_{T\gamma V}\left(8 \sin \theta_T +11\sqrt{2} \cos \theta_T\right)
\frac{2\sqrt{3} F_V}{27}\left(D^{-1}_{M_V}(q_1^2)+D^{-1}_{M_V}(q_2^2)\right)
\\
&+C_{TVV}\left( \sin \theta_T +2\sqrt{2} \cos \theta_T\right)
\frac{2\sqrt{2} F_V^2}{3\sqrt{3}} D^{-1}_{M_V}(q_1^2)D^{-1}_{M_V}(q_2^2)
\\
&+C_{TVV^\prime}\left( \sin \theta_T +2\sqrt{2} \cos \theta_T\right)
\frac{2\sqrt{2} F_V F_{V^\prime}}{3\sqrt{3}}
D^{-1}_{M_V}(q_1^2)D^{-1}_{M_{V^\prime}}(q_2^2)
+V\leftrightarrow V^\prime\Bigg],
\end{split}
\end{equation}
\begin{equation}
\begin{split}
\mathcal{F}_1^{f^\prime_2}(q_1^2,q_2^2)=\mathcal{F}_1^{f_2}(q_1^2,q_2^2)|_{\sin \theta_T\to \cos \theta_T,\cos\theta_T\to-\sin\theta_T, M_{f_2} \to M_{f^\prime_2}}\,,
\end{split}
\end{equation}
where $D_{M_R}(q_i^2)=M_R^2-q_i^2$. The $\mathcal{F}_3^T$ FF is obtained straightforwardly, noting that
\begin{equation}
    \mathcal{F}_3^T(q_1^2,q_2^2)=M_T^2 \mathcal{F}_1^T(q_1^2,q_2^2)|_{C_{T}\to g_{T}}.
\end{equation}

Ref.~\cite{Hoferichter:2020lap} derived the behavior of the $\mathcal{F}_i^T$ form factors in the asymptotic region, that will be discussed next in the relevant $i=1,3$ cases. 
\subsection{$\mathcal{F}_1^T$ high-energy behavior}
Since in a model with finite resonances only a discrete amount of short-distance constraints (SDCs) can be imposed, for the $\mathcal{F}_1^T$ FF we choose to fix the symmetric double virtual case and the single virtual $1/Q^4$ behavior
\begin{subequations}
\label{eq:TFF1SDC}
\begin{equation}
    \lim_{Q^2 \to \infty} \mathcal{F}_1^T(-Q^2,-Q^2)=-\frac{3F_T^{\rm eff}M_T^3}{14Q^4},
    \label{eq:DVSDC}
\end{equation}
\begin{equation}
    \lim_{Q^2 \to \infty}\mathcal{F}_1^T(-Q^2,0)\sim \frac{1}{Q^4},
    \label{eq:SVSDC}
\end{equation}
\label{SDC}
\end{subequations}
where $Q_i^2=-q_i^2$, and $F_{f_2}^{\rm eff}=4\sum_a C_a F_{f_2}^a=79(8)$ MeV~\cite{Hoferichter:2020lap}. The values of this effective constant for $a_2,f_2'$ are obtained from the relation $\frac{F_T^{\rm eff}}{c_T}=\frac{F_{f_2}^{\rm eff}}{c_{f_2}}$, where $c_T$ is a flavor space coefficient, given by: $c_{a_2}=1,$ $ c_{f_2}=\frac{\sin \theta_T+2\sqrt{2}\cos \theta_T}{\sqrt{3}},$ $ c_{f^\prime_2}=\frac{\cos \theta_T-2\sqrt{2}\sin \theta_T}{\sqrt{3}}$. We note that the magnitude of $\mathcal{F}_1^T$ is overestimated this way, since for one finite and another asymptotic photon virtuality, $\mathcal{F}_1^T$ should vanish as $\sim \ln Q^2/Q^6$ that we will approximate as $\sim 1/Q^4$ because we are working with a rational approximant that cannot produce the logarithmic term with a finite number of multiplets). However, with two multiplets of vector meson resonances (and also with three of them), we get $\mathcal{F}_1^T\sim 1/Q^2$ in this mixed region, which we consider as a systematic error when computing $a_\mu^{\rm HLbL:\, T-pole}$. 
The resulting FFs, after imposing the high-energy constraints mentioned above are: 
\begin{equation}
    \mathcal{F}_1^T(q_1^2,q_2^2)=c_T M_T\frac{9\,\tilde F_T^{\rm eff}\tilde M_T^2(M_V^4-q_1^2 q_2^2)/\tilde c_T-14\sqrt{2}C_{TVV}F_V^2 (M_V^2-M_{V^\prime}^2)^2}{42\,D_{M_V}(q_1^2)D_{M_V}(q_2^2)D_{M_{V^\prime}}(q_1^2)D_{M_{V^\prime}}(q_2^2)}.
    \label{eq:ConstrainedTFF1}
\end{equation}
We cannot impose the SDCs to all tensor mesons individually, and we have to choose one, with mass $\tilde M_T$, effective decay constant $\tilde F_T^{\rm eff}$ and flavor space coefficient $\tilde c_T$. From the constraints from eq. (\ref{eq:TFF1SDC}), 5 relations between couplings are found~\footnote{We have chosen $a_2$, since its mass and flavor space factor are closer to the isospin average of the three, which minimizes the systematic error induced by this choice. This will be discussed in the section on systematic uncertainties.}. The only free parameters left are the combination $C_{TVV}F_V^2$ and the mixing angle $\theta_T$. These two parameters can be used to normalize the form factors to reproduce the radiative decay widths as was done in ref.~\cite{Chen:2023ybr}. We will take their chiral limit value for the mixing angle, $\theta_T=(27.2\pm1.1)^\circ$, and fix $C_{TVV}F_V^2=(0.110\pm0.005)$ GeV$^3$, from the measured value of $\Gamma(a_2\to\gamma \gamma)$. Given the SDCs, it is also possible to evaluate the other couplings, namely: $C_{TVV^\prime}F_V F_{V^\prime}=(-0.116\pm 0.005)\,\rm{GeV}^3$ and $C_{TV^\prime V^\prime}F_{V^\prime}^2=(0.158\pm 0.007)\,\rm{GeV}^3$. The corresponding uncertainties are negligible with respect to the one obtained in the next section.
\subsection{$\mathcal{F}_3^T$ high-energy behavior}
For the $\mathcal{F}_3^T$ FF, 
we have chosen to fix the doubly virtual behavior, both symmetric and asymmetric, to their leading power law:
\begin{subequations}
\begin{equation}
    \lim_{Q^2\to\infty}\mathcal{F}_3^T(-Q^2,-Q^2)=-\frac{8}{21}\frac{F_T^{\rm eff}M_T^5}{Q^6}\,,
\end{equation}
\begin{equation}
    \lim_{Q^2\to\infty}\mathcal{F}_3^{T}(-Q^2,-\lambda Q^2)\sim \frac{1}{Q^6},\quad \lambda \in (0,1)
    \,.
\end{equation}
\end{subequations}
For the symmetric doubly virtual case, we have matched the coefficient to the one given by the Light-Cone-Expansion in ref. \cite{Hoferichter:2020lap}. As in $\mathcal{F}_1^T$, we had to choose a tensor meson TFF to impose the SDC. For the same reasons discussed before, we have chosen $a_2$ also in this case.

\begin{figure}
    \centering
    \includegraphics[width=1\linewidth]{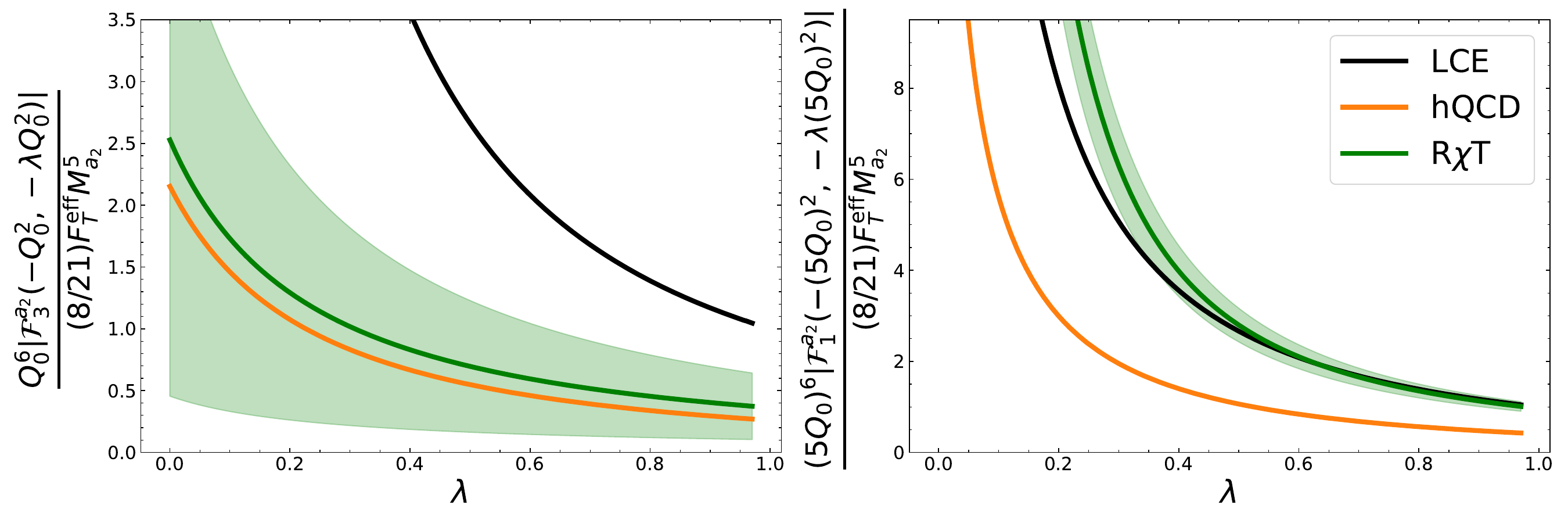}
    \caption{Transition from the purely hadronic to the asymptotic region as a function of the relative virtuality of both photons for $\mathcal F_3^T$, $\lambda$, at the matching scale $Q_0=1.5\,\rm{GeV}$(left) and $5Q_0$(right) for the 3 models considered and the Light-Cone-Expansion. Our one $\sigma$ uncertainties are displayed by the green band.}
    \label{fig:matching3}
\end{figure}

Imposing the SDCs results in~\footnote{The aforementioned behavior is achieved for all asymmetries except $\lambda=0$. In this case, the symmetric part of the SDCs in ref. \cite{Hoferichter:2020lap} behaves as $\frac{1}{Q^6}$, but in our approach it goes as $\frac{1}{Q^2}$. We account for this as a systematic uncertainty.} 

\begin{equation}
\mathcal{F}_3^T(q_1^2,q_2^2)= \frac{c_T}{\tilde c_T} \left(\frac{M_T}{\tilde M_T}\right)^3\frac{\frac{4}{21} \tilde F_{T}^{\rm eff}\tilde M_T^5\left(q_1^2+q_2^2\right)+\tilde{\mathcal{F}}_3^T(0,0) M_V^4 M_{V^\prime}^4 }{D_{M_V}(q_1^2)D_{M_V}(q_2^2)D_{M_{V^\prime}}(q_1^2)D_{M_{V^\prime}}(q_2^2)},
\label{eq:F3}
\end{equation}
where the normalization $\tilde{\mathcal{F}}_3^T(0,0)$ is left free-- with the same notation as in \ref{eq:ConstrainedTFF1}--, for future fits in the advent of data. 
A reasonable ballpark of the value of this form factor at zero virtuality can be obtained within $R \chi T$. This can be done by requiring that at some scale $Q_{\rm asympt}$, the transition of $\mathcal{F}_T^3$ from our model to the asymptotic behavior of the Light-Cone-Expansion --parametrized by $\frac{\mathcal{F}_T^{3\,\rm{R\chi T}}(-Q_{\rm asympt}^2,-Q^2_{\rm asympt})}{F^{3\,\rm{LCE}}_T(-Q_{\rm asympt}^2,-Q^2_{\rm asympt})}$-- matches the one of $\mathcal{F}_T^1$ within $1\sigma$ for $\lambda=1$ (this procedure is detailed in the \nameref{sec:appendix}). For $\mathcal{F}_T^1$, this transition is smooth at $Q_{\rm asympt}=5Q_0$ --as it can be seen in Figures \ref{fig:matching1} and \ref{fig:matching3}--, where $Q_0$ is the upper limit of the purely hadronic region, 1.5 GeV.\\
This results in the values $\mathcal{F}_{a_2}^3(0,0)=-(0.106\pm0.101)$, $\mathcal{F}_{f_2}^3(0,0)=-(0.165\pm0.158)$, $\mathcal{F}_{f^\prime_2}^3(0,0)=-(0.038\pm0.036)$; and given the SDCs, we get the couplings $g_{TVV}F_V^2 =(0.044\pm 0.047 )\rm{GeV}$, $g_{TVV^\prime}F_V F_{V^\prime}=-(0.032\pm0.049 )\rm{GeV}$ and $g_{TV^\prime V^\prime}F_{V^\prime}^2=(0.021\pm0.051 )\rm{GeV}$.\\

\section{Form Factors and $a_\mu$ computation}\label{sec:4}
\begin{figure}[h!]
    \centering
    \includegraphics[width=1\linewidth]{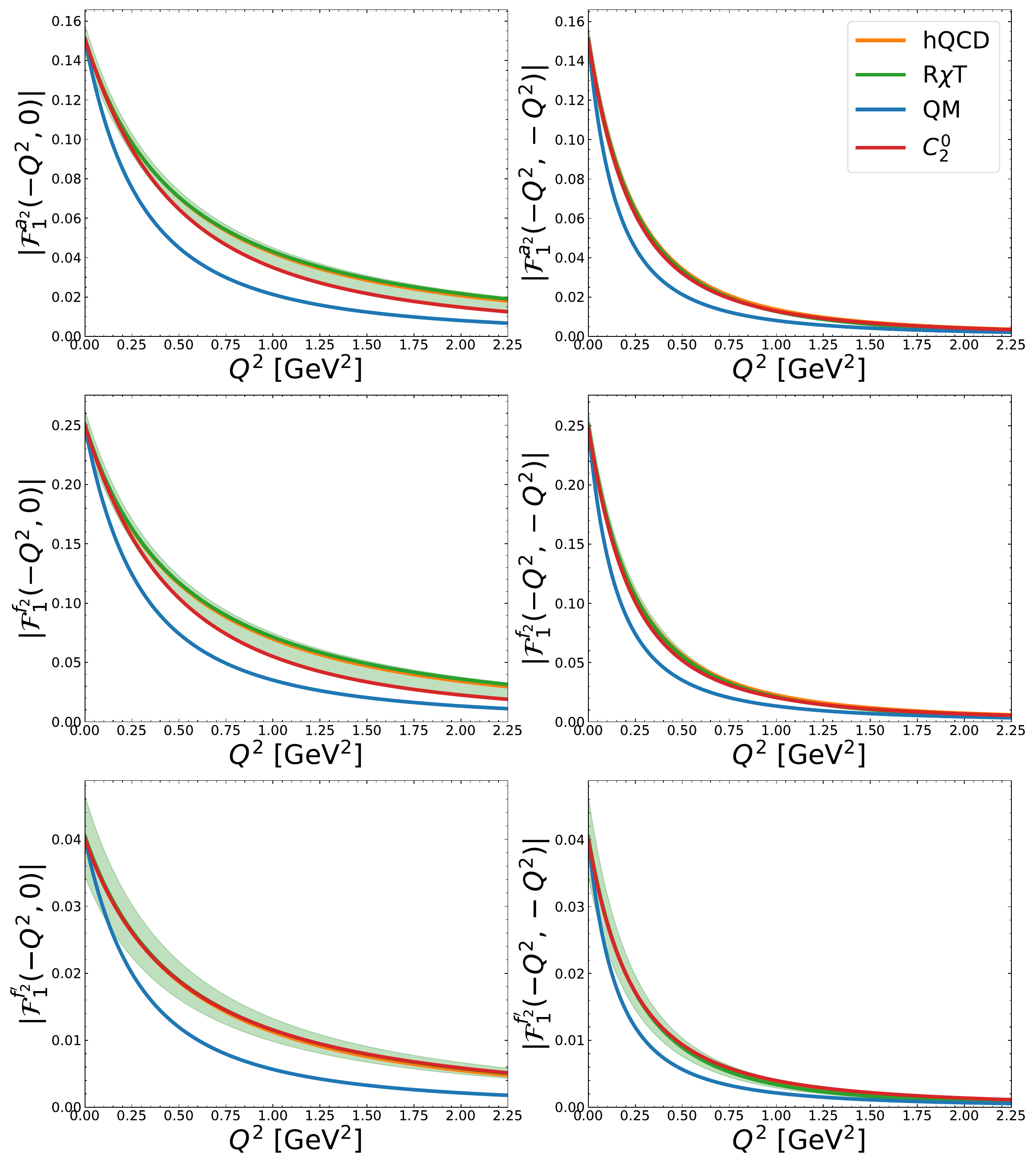}
    \caption{Comparison between the simple quark model (QM) \cite{Schuler:1997yw} used in the dispersive determination \cite{Hoferichter:2024bae}, the holographic QCD (hQCD) hard-wall model~\cite{Cappiello:2025fyf}, the $R \chi T$ model used in this work and the $C_2^0$ of eq. (\ref{eq_defCA}) for the single and symmetric double virtual form factor $\mathcal{F}_1^{T}$, for all 3 tensor mesons: $a_2$(1320), $f_2$(1270) and $f_2^\prime$(1525). Our one $\sigma$ uncertainties are displayed by the green band.}
    \label{fig:DispvsRchiT}
\end{figure}

\begin{figure}[h!]
    \centering
    \includegraphics[width=1\linewidth]{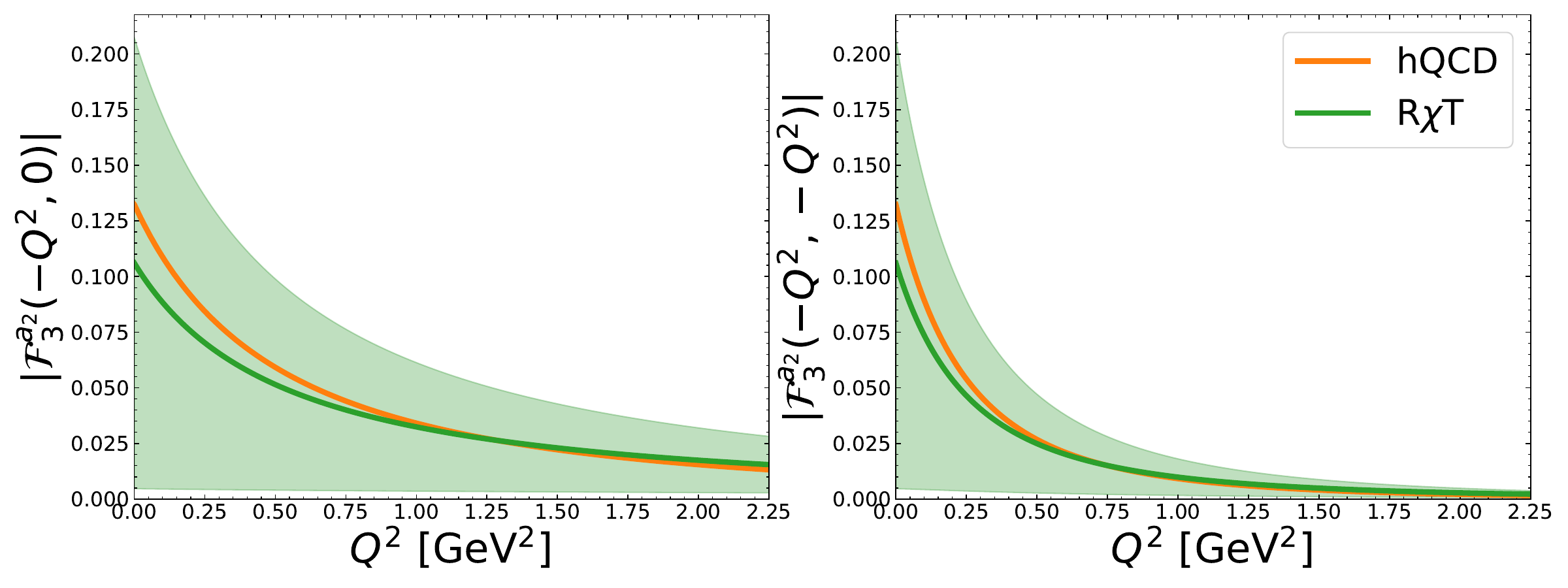}
    \caption{Comparison between the (hQCD) hard-wall model~\cite{Cappiello:2025fyf} and the $R \chi T$ model used in this work for the single and symmetric double virtual form factor $\mathcal{F}_3^{T}$, for $a_2$(1320). The same shape is found for $f_2,f_2'$ (including uncertainties), with a relative factor of $c_T \frac{M_T^3}{M_a^3}$ with respect to the one shown. Our one $\sigma$ uncertainties are displayed by the green band, which, in contrast to $\mathcal F_1^T$, is dominated by the large error of $\mathcal{F}_3^T(0,0)$}.
    \label{fig:hQCDvsRchiTF3}
\end{figure}

For the rest of 
parameters 
in eqs. (\ref{eq:ConstrainedTFF1}) and (\ref{eq:F3}), we use the PDG  input~\cite{ParticleDataGroup:2024cfk} for the tensor meson masses and their radiative decay widths. The effective masses of the vector meson multiplets are obtained from the study of $a_\mu^{\rm P-pole}$, within $R\chi T$, in  ref.~\cite{Estrada:2024cfy} and $F_T^{\rm eff}$ is taken from refs.~\cite{Hoferichter:2020lap, Braun:2016tsk}. Thus, with all parameters fixed -within uncertainties- we complete our description of the tensor mesons $\mathcal{F}_1^T$ and $\mathcal{F}_3^T$, which are shown in Figures \ref{fig:DispvsRchiT} and \ref{fig:hQCDvsRchiTF3} for the single virtual and the symmetric double virtual case of $a_2,\, f_2 \,\rm{and}\, f_2^\prime$. For $\mathcal F_{1}^T$ all three are shown, since they provide complementary information; while for $\mathcal{F}_3^T$, only $a_2$ is displayed, since their shapes are similar and their large uncertainty is saturated by that of $\mathcal F_3^T(0,0)$. 

There is scant related experimental information, Belle \cite{Belle:2015oin} measured the single virtual form factors for the $f_2(1270)$ tensor meson in the helicity basis. The needed change of basis is \cite{Hoferichter:2020lap}:
\begin{subequations}
\begin{equation}
    \mathcal{F}_{\lambda=0}^T=\frac{Q^2}{\sqrt{6}M_T^2}\mathcal{F}_1^T(-Q^2,0)-\frac{(M_T^2+Q^2)^2}{2\sqrt{6}M_T^4}\mathcal{F}_2^T(-Q^2,0)+\frac{Q^2}{\sqrt{6}M_T^2}\mathcal{F}_5^T(-Q^2,0),
\end{equation}
\begin{equation}
    \mathcal{F}_{\lambda=1}^T=\frac{\sqrt{Q^2}}{\sqrt{2}M_T}\mathcal{F}_1^T(-Q^2,0)+\frac{\sqrt{Q^2}(M_T^2-Q^2)}{2\sqrt{2}M_T^3}\mathcal{F}_5^T(-Q^2,0),
\end{equation}
\begin{equation}
    \mathcal{F}_{\lambda=2}^T=-\mathcal{F}_1^T(-Q^2,0)+\frac{Q^2}{M_T^2}\mathcal{F}_5^T(-Q^2,0).
\end{equation}
\label{eq:helicitybasis}
\end{subequations}
In our case, this is simplified, as $\mathcal{F}^T_{2,5}(-Q^2,0)=0$. The comparison of these data with the quark model used in \cite{Hoferichter:2024bae}, the Hard-Wall model used in \cite{Cappiello:2025fyf}, and our results is shown in Figure \ref{fig:BellevsTh}~\footnote{Of particular interest is the first point of the upper left plot, since it can only be explained with a non-zero $\mathcal{F}_2^T$, according to eqs. (\ref{eq:helicitybasis}).}. Our model exhibits good agreement with the experimental data. Unfortunately, this cannot be performed for $\mathcal{F}_3^T$, since there are no measurements probing it. However, 
this is an interesting direction of future research work, once double virtuality data are obtained.
\begin{figure}[h!]
\centering
\includesvg[width=\textwidth]{FLambdaF2Comparisson.svg}
\caption{Comparison between the simple quark model (QM)~\cite{Schuler:1997yw} used in ref.~\cite{Hoferichter:2024bae}, the hard-wall model (hQCD) \cite{Cappiello:2025fyf} and this work ($R\chi T$) with the Belle data \cite{Belle:2015oin} in the helicity basis of eq. (\ref{eq:helicitybasis}) for the $f_2$(1270) tensor meson, normalized by $\mathcal{F}^{f_2}(0,0)=\sqrt{\frac{5 \Gamma_{T\gamma\gamma}}{\pi \alpha^2 M_T}}$. Our one $\sigma$ uncertainties are displayed by the green band.}
    \label{fig:BellevsTh}
\end{figure}

It is important to remark that --by construction-- these TFFs fulfill the doubly virtual asymptotic behavior given in ref.~\cite{Hoferichter:2020lap}. However, for arbitrary asymmetries --parametrized by $w=\frac{q_1^2-q_2^2}{q_1^2+q_2^2}$, so that 
$\mathcal{F}_i^T(q_1^2,q_2^2)=\mathcal{F}_i^T(\tilde Q^2)f_i^T(w)$--, for $\tilde Q^2=\frac{q_1^2+q_2^2}{2}\sim \infty$~\footnote{Notice that this notation for $Q^2$ is used in eq. (\ref{eq:RchiTHighEnergy}) as in \cite{Hoferichter:2020lap}, which differs from the $Q_i^2=-q_i^2$ that we have used before in this work. Particularly, for the symmetric double virtual case in the spacelike region, they have the same magnitude but different signs.}, we get:
\begin{subequations}
    \begin{equation}
    \mathcal{F}_1^T(q_1^2,q_2^2) \to \frac{4 \sum_a C_a F_T^a m_T^3}{{\tilde Q^4}} \left(-\frac{3}{14}(1-w^2)^{-1}\right),
    \end{equation}
    \begin{equation}
     \mathcal{F}_3^T(q_1^2,q_2^2)\to \frac{4 \sum_a C_a F_T^a m_T^5}{\tilde Q^6}
     \left(\frac{8}{21}(1-w^2)^{-2}\right),
    \end{equation}
    \label{eq:RchiTHighEnergy}
\end{subequations}
which is simpler than the ones obtained by refs.~\cite{Hoferichter:2020lap} and \cite{Cappiello:2025fyf}. One reason for the lack of structure of our result is the truncation of the infinite tower of interactions to two vector and the lightest tensor meson multiplets. If the infinite tower of remaining resonances were not integrated out, a logarithmic term would be expected, as in the quoted refs. 

\begin{figure}
    \centering
    \includegraphics[width=1\linewidth]{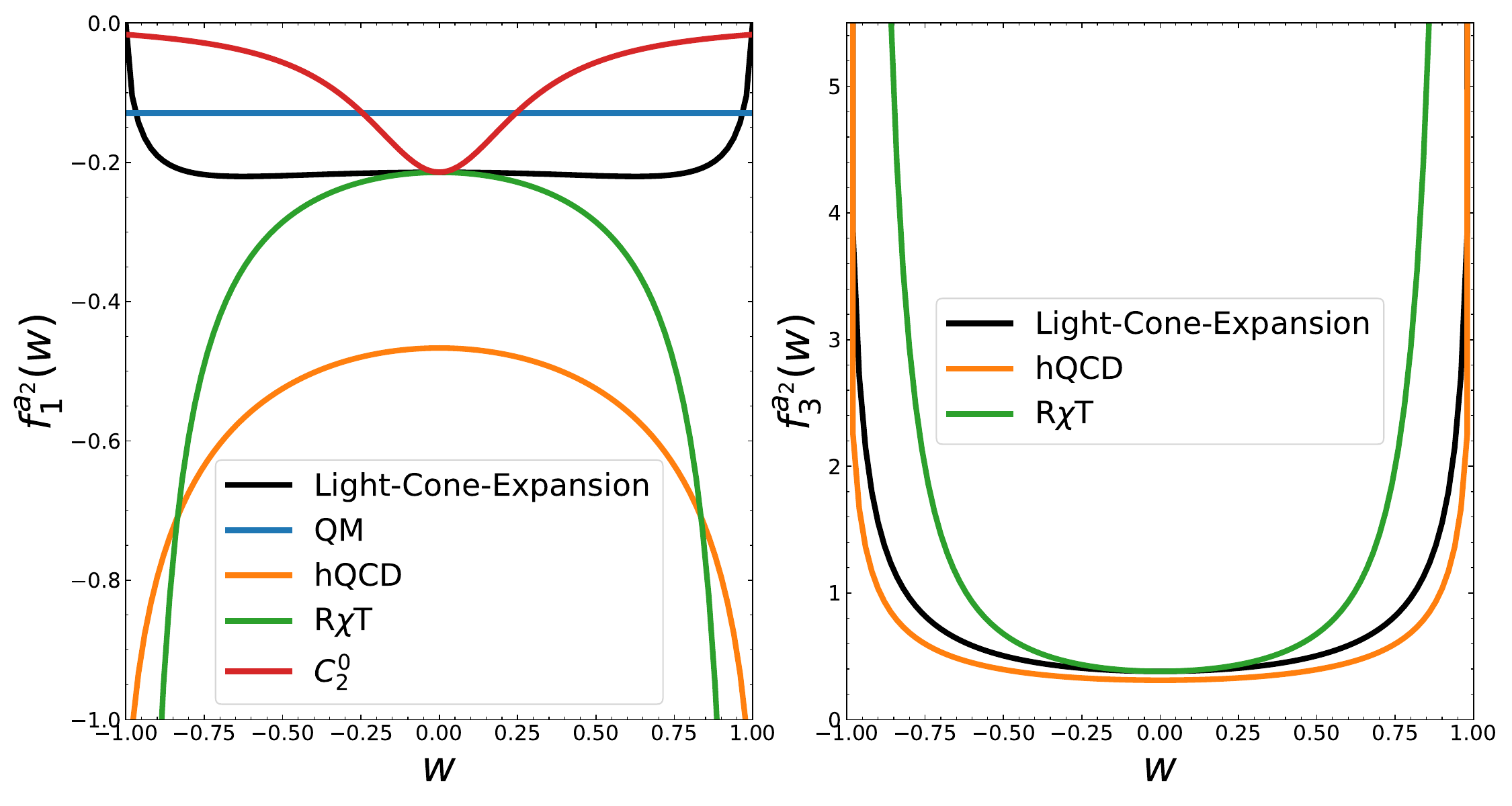}
    \caption{Comparison of the full asymmetry range of the asymptotic behavior of $f_1^{a_2}(w)$ given by ref.~\cite{Hoferichter:2020lap} using the Light-Cone-Expansion, the quark model (QM) \cite{Schuler:1997yw} used in ref.~\cite{Hoferichter:2024bae}, the hard-wall model of ref.~\cite{Cappiello:2025fyf} (hQCD), this work ($R\chi T$)  (left plot), and the Canterbury Approximant used to correct our asymptotic behavior ($C_2^0$). The same comparison for Light-Cone-Expansion, hQCD and this work 
    is shown in the right plot  for $f_3^{a_2}(w)$.}
    \label{fig:asymmetry_comparisonf1f3}
\end{figure}

The comparison between the asymmetry functions for the $a_2$ --which is representative for all tensor mesons--, is shown in Figure \ref{fig:asymmetry_comparisonf1f3}, where it is clear that none of the 3 models that have been used for computing the $a_\mu^{\rm HLbL:T-poles}$ fulfills the asymmetric asymptotic behavior predicted by ref.~\cite{Hoferichter:2020lap} for $f_1^T$, being generally worse as the asymmetry approaches the single virtual case. For the $f_3^T$, the asymptotic behavior fails to be reproduced in the asymmetric part, since the logarithmic divergence at $w\to\pm1$ cannot be obtained in a finite resonance model. Given the relevant region for $a_\mu$, the second one was prioritized.  
We note that the sign of $\mathcal{F}_3^T/\mathcal{F}_1^T$ at high energies agrees in all approaches: Light-Cone-Expansion\cite{Hoferichter:2020lap}, Hard-Wall Model\cite{Cappiello:2025fyf}, and this work.

\begin{figure}
    \centering
    \includegraphics[width=1\linewidth]{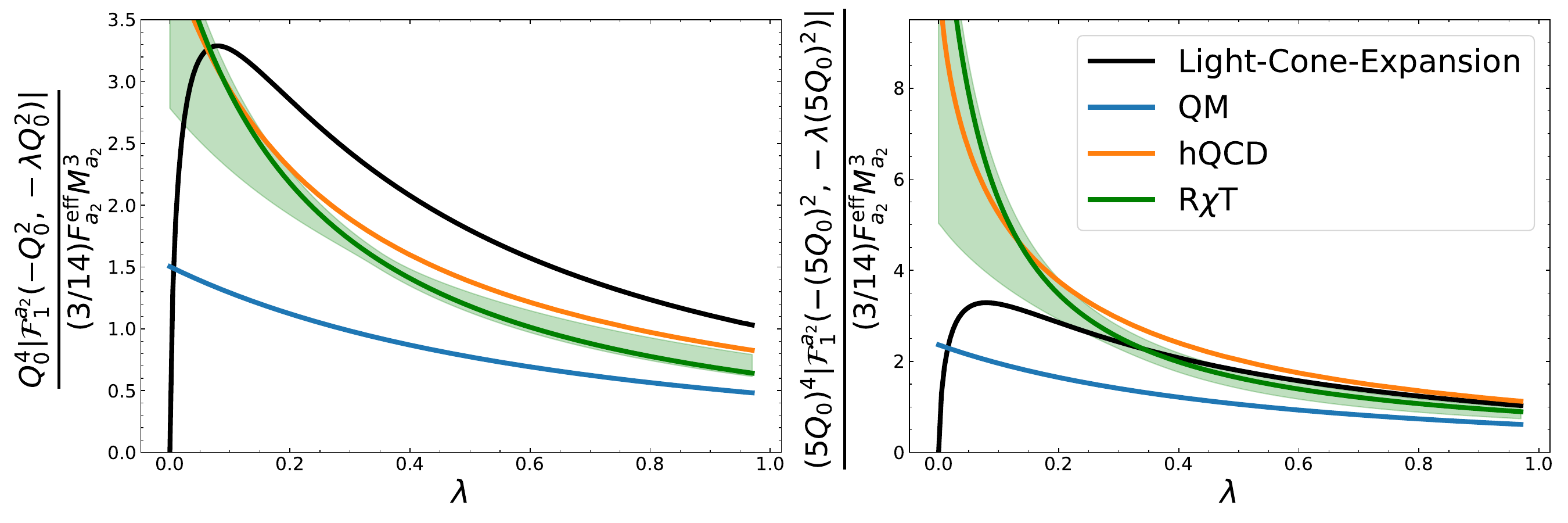}
    \caption{Transition from the purely hadronic to the asymptotic region as a function of the relative virtuality of both photons for $\mathcal F_1^T$, $\lambda$, at the matching scale $Q_0=1.5\,\rm{GeV}$(left) and $5Q_0$(right) for the 3 models considered and the Light-Cone-Expansion. Our one $\sigma$ uncertainties are displayed by the green band.}
    \label{fig:matching1}
\end{figure}

However, noticeably, in the purely hadronic region where these contributions are computed, the high-energy behavior does not significantly influence the resulting contributions to $a_\mu$
. The matching at $Q_0=1.5$ GeV is shown in the left plot of Figure \ref{fig:matching1} for $\mathcal{F}_1^T$. 
There it can be seen, again, that the single virtual asymptotic behavior ($\lambda=0$) is not reproduced
. Near the symmetric ($\lambda=1$) asymptotic behavior, a better transition is achieved --as expected-- for higher values of the matching scale $Q_0=7.5\, \rm{GeV}$, as it can be appreciated in the right plot of Figure \ref{fig:matching1}.

For the $a_\mu$ computation, we use the master formula provided in ref.~\cite{Colangelo:2015ama}, using the optimized basis in ref.~\cite{Hoferichter:2024fsj}, which allows an evaluation of the tensor meson poles without kinematic singularities for specific cases, including the one where only $\mathcal{F}_1^T$ is non-zero and the more general case where both $\mathcal{F}_1^T$ and $\mathcal{F}_3^T$ are non-vanishing.
In this way, we got the following results (in units of $10^{-11}$):\\

$\bullet$ Only with $\mathcal{F}_1^T$
\begin{subequations}\label{eq_T-poles}
    \begin{equation}
        a_\mu^{\rm a_2-pole}=-\left(1.02(10)_{\rm stat}(^{+0.00}_{-0.12})_{\rm syst}\right),
        \end{equation}
    \begin{equation}
        a_\mu^{\rm f_2-pole}=-\left(3.2(3)_{\rm stat}(^{+0.0}_{-0.4})_{\rm syst}\right),
    \end{equation} 
    \begin{equation}\label{eq_T-polesf2p}
        a_\mu^{\rm f_2^\prime-pole}=-\left(0.042(13)_{\rm stat}\left(^{+0.000}_{-0.008}\right)_{\rm syst}\right),
    \end{equation}
    \begin{equation}
        a_\mu^{a_2+f_2+f_2^\prime \rm - pole}=-\left(4.3^{+0.3}_{-0.5}\right).
    \end{equation}
\end{subequations}
\\
$\bullet$ With both $\mathcal{F}_1^T$ and $\mathcal{F}_3^T$ \
\begin{subequations}\label{eq_T-poles_bothFFs}
    \begin{equation}
        a_\mu^{\rm a_2-pole}=+0.47(1.43)_{\rm norm}(3)_{\rm stat}(^{+0.06}_{-0.00})_{\rm syst}, 
        \end{equation}
    \begin{equation}   
        a_\mu^{\rm f_2-pole}=+1.18(4.18)_{\rm norm}(12)_{\rm stat}(^{+0.24}_{-0.00})_{\rm syst},
    \end{equation} 
    \begin{equation}
    \label{eq_T-poles_bothFFsf2p}
        a_\mu^{\rm f_2^\prime-pole}=+0.040(78)_{\rm norm}(2)_{\rm stat},
    \end{equation}
    \begin{equation}
        a_\mu^{a_2+f_2+f_2^\prime \rm - pole}=+1.7 (4.4) ,
    \end{equation}
\end{subequations}
where we have reported the errors coming from the lack of information on $\mathcal{F}_3^T(0,0)$--which we call norm in eq. (\ref{eq_T-poles_bothFFs})--, the systematics and the statistics, which are detailed next.
\subsection{Uncertainty Computation}
The statistical errors come from the propagation of the uncertainties on all parameters used to compute the $a_\mu$ contributions, whereas the systematic ones come --in principle-- from two main sources for $\mathcal{F}_1^T$: first, concerning the tensor meson $T$ chosen to impose eq. (\ref{eq:DVSDC}); second, fulfilling all known SDCs being impossible in our setting. However, our main source of error is the lack of information on the value of $\mathcal{F}_T^3$ at zero virtuality. As discussed at the end of section 3, we floated this value, matching it with the high-energy behavior of $\mathcal{F}_T^1$ at different scales, leading to a large error, which is expected to be sharply reduced and become of statistical origin when it be measured.
\subsubsection{Choice of tensor meson's SDCs}
The short-distance limits of the 
$\mathcal{F}_1^T$ FF can be imposed only to one of the three particles (either $a_2$, $f_2$ or $f_2^\prime$), which causes the modification $M_T^3\to M_T \tilde M_T^2$ in the asymptotic term of eq. (\ref{eq:ConstrainedTFF1}). We used the respective mass for each tensor meson instead of choosing a common one for all 3. To compute the systematic error induced by this choice, we varied this parameter ($\tilde M_T$) among 4 options: the 3 particle masses and their isospin average. This error resulted negligible compared to the remaining systematics for this form factor, and even more with the statistical uncertainties. For $\mathcal{F}_3^T$, it is further suppressed by the one coming from $\mathcal F_3^T(0,0)$. 
\subsubsection{High-energy behavior}
Our $R\chi T$ result for the $\mathcal{F}_1^T$ FF
, eq. (\ref{eq:ConstrainedTFF1}), does fulfill the two constraints from eq. (\ref{SDC}); however, the limiting case when only one of the $Q_i$ is asymptotic and the other one is finite (but non-vanishing), should scale as $\sim \ln Q^2/Q^6$, while it goes as $1/Q^2$. To estimate how this mismatch induces an error in our computation of $a_\mu^{\rm HLbL:\, T-pole}$, we construct a Canterbury Approximant in the same fashion as in ref.~\cite{Masjuan:2017tvw}. We will impose a $1/Q^4$ behavior, since it is the best one can do with a rational approach to improve the $1/Q^2$ behavior. For this purpose, the minimal choice is a $C_2^0(Q_1^2,Q_2^2)$:
\begin{equation}\label{eq_defCA}
    \mathcal{F}_1^T(Q_1^2,Q_2^2)=\frac{\mathcal{F}_1^T(0,0)}{1+\beta^T_1(Q_1^2+Q_2^2)+\beta^T_{1,1}Q_1^2 Q_2^2+\beta^T_2 (Q_1^4+Q_2^4)},
\end{equation}
where we omit the terms with $\beta^T_{1,2}$ and $\beta^T_{2,2}$ coefficients, as they contribute to orders $1/Q^6$ and $1/Q^8$ in the  asymptotic behavior. All single virtual, double virtual, and mixed asymptotic behaviors can be reproduced by the condition:
\begin{equation}
    \beta^T_2=\frac{{-}14 \mathcal{F}_1^T(0,0)-3 \beta_{1,1}F_T^{\rm eff}M_T^3}{6 F_T^{\rm eff}M_T^3}.
\end{equation}

The 2 remaining constants, $\beta_1^T$ and $\beta_{1,1}^T$, will be fixed by matching the first two low-energy-expansion constants from our model and the considered $C_2^0(Q_1^2,Q_2^2)$, eq.~(\ref{eq_defCA}):
\begin{equation}
    \frac{\mathcal{F}_{1}^T(Q_1^2,Q_2^2)}{\mathcal{F}_1^T(0,0)}\approx 1-\frac{a_1^{T}}{M_T^2}(Q_1^2+Q_2^2)+\frac{b^T_{1,1}}{M_T^4}Q_1^2Q_2^2+..., 
\end{equation}
leading to the following relations:
\begin{subequations}
    \begin{equation}
        \beta^T_1=\frac{1}{M_\Lambda^2}=\frac{1}{M_V^2}+\frac{1}{M_{V^\prime}^2},
    \end{equation}
    \begin{equation}
        \beta^T_{1,1}=\frac{1}{M_{\Lambda^\prime}^4}=\frac{14\sqrt{2}C_{TVV}F_V^2(M_V^4-M_{V^\prime}^4)^2-9\tilde F_T^{\rm eff}\tilde M_T^2M_V^4(M_V^4+2M_V^2M_{V^\prime}^2+2M_{V^\prime}^4)/\tilde c_T}{M_V^4 M_{V^\prime}^4\left[14\sqrt{2 }C_{TVV}F_V^2(M_V^2-M_{V^\prime}^2)^2-9 \tilde F_T^{\rm eff}\tilde M_T^2M_V^4/\tilde c_T\right]},
    \end{equation}
\end{subequations}
where we introduced the two scales $M_\Lambda\sim M_{\Lambda^\prime}\sim M_\rho$. The high-energy behavior {($\tilde Q^2\sim \infty $)} of the rational approximant form factor is given by:
\begin{equation}
     \mathcal F_1^{T\rm{-C_2^0}}(q_1^2,q_2^2)\to \frac{4\sum_a C_a F_T^a M_T^3}{\tilde Q^4}\left( -\frac{3\mathcal F_1^T(0,0) M_{\Lambda^\prime}^4}{14\mathcal F_1^T(0,0)M_{\Lambda^\prime}^4(w^4+6w^2+1)+24 F_{T}^{\rm eff} M_T^3 w^2}\right).
\end{equation}
The improved high-energy behaviour achieved with this Canterbury Approximant can be appreciated in Figure \ref{fig:asymmetry_comparisonf1f3}.
We compute the associated error by comparing the $a_\mu^{\rm T-pole}$ value of our model to the results obtained in the $C_2^0$ model, where both the asymptotic behavior and the normalization are improved with respect to $R \chi T$. This indicates that --for the original setting-- the magnitude of $a_\mu^{T\rm-pole}$ is overestimated for the $a_2$ and $f_2$ and negligible for $f_2^\prime$, which is the reason why the result from eq. (\ref{eq_T-polesf2p}) does not display this error.


Naturally, our result neglecting $\mathcal{F}_3^T$, $a_\mu^{a_2+f_2+f_2^\prime
}=$ $-\left(4.3^{+0.3}_{-0.5}\right) \times 10^{-11}$, agrees in sign with the dispersive computation, $a_\mu^{\rm a_2+f_2+f_2^\prime}=-2.5(3)\times10^{-11}$ \cite{Hoferichter:2024bae}. A comparison of our $\mathcal{F}_1^T$ in $R\chi T$ with the one of the simple quark model\cite{Schuler:1997yw} used in the dispersive determination \cite{Hoferichter:2024bae}, Figure \ref{fig:DispvsRchiT}, shows that in the purely hadronic region ($Q_i<1.5\, \rm{GeV}$)~\footnote{The contribution to $a_\mu^{\rm T-pole}$ is evaluated in this region, as it has become standard in the matching to the OPE results~\cite{Vainshtein:2002nv,Melnikov:2003xd,Knecht:2003xy,Bijnens:2019ghy,Bijnens:2020xnl,Bijnens:2021jqo,Bijnens:2022itw,Bijnens:2024jgh}.} the TFF from $R \chi T$ is larger than the quark model one, which explains why we get a larger magnitude for $a_\mu$. Noteworthy, they are in very nice agreement with the holographic ones, once the $\mathcal{F}_3^T$ contribution (responsible for the sign change) has been discarded, $a_\mu^{\rm a_2+f_2+f_2^\prime}=-4.5(6)\times10^{-11}$~\cite{Cappiello:2025fyf} --which is clearly explained by the agreement of the $\mathcal{F}_1^T$ between these two models in Figure \ref{fig:DispvsRchiT} at both single virtual and symmetrically double virtual cases--.

This picture changes, however, once we include the $\mathcal{F}_3^T$ contribution. In this case, not only is the mismatch in the asymptotic behavior of $\mathcal{F}_1^T$ --as discussed at the beginning of this subsection-- relevant, but also the deviant $1/Q^2$ behavior of the symmetric part of the single virtual $\mathcal{F}_3^T$ in the $Q^2\to\infty$ limit. By considering the same Canterbury Approximant as above, $C_{2}^0(Q_1^2, Q_2^2)$, but restoring $\beta^T_{1,2}$, an $\mathcal{O}(1/Q^6)$ damping can be obtained; however, this results in negligible improvement of our $R \chi T$ result due the lack of information on $\mathcal F_3^T(0,0)$. The systematics related to the wrong asymptotic behavior for $\mathcal{F}_1^T$ are included for consistency with the results in eq. (\ref{eq_T-poles}). They are dominant with respect to the rest of the systematics~\footnote{Estimating the uncertainty caused by cutting the resonance spectrum by including an additional multiplet of vector resonances leads to a much smaller uncertainty, for instance.}, but strongly suppressed with respect to the $\mathcal F_3^T(0,0)$ uncertainty.

\section{Conclusions}\label{sec:Concl}
The FNAL Muon g-2 Collaboration will soon release its final measurement of this precision observable, with potentially crucial beyond the Standard Model implications. This has triggered and boosted the efforts of the Muon \ensuremath{g - 2} Theory Initiative, towards reducing the uncertainty of the Standard Model prediction, aiming to maximize the reach on new physics of this measurement. Within this global effort, not only the leading uncertainties --of hadronic origin, coming from the vacuum polarization piece--, but also the subleading ones, coming from the light-by-light part, have been improved recently.\\

In this work we have focused on the tensor-pole contribution, in which the most recent evaluations, dispersive (with a quark model FF) and holographic, differ in sign. We have studied it, within Resonance Chiral Theory
, first in a simplified consistent setting, where only one form factor ($\mathcal{F}_1^T$) is non-vanishing, as in the quark model approach. This makes this contribution negative (the $\mathcal{F}_3^T$ contribution is responsible for its positiveness in holographic QCD). We have compared our $\mathcal{F}_1^T$ to both computations and found closer agreement with the holographic form factor. Our results (in units of $10^{-11}$): 
$a_\mu^{\rm a_2-pole}=-\left(1.02(10)_{\rm stat}(^{+0.00}_{-0.12})_{\rm syst}\right)$, $a_\mu^{\rm f_2-pole}=-\left(3.2(3)_{\rm stat}(^{+0.0}_{-0.4})_{\rm syst}\right)$ and $a_\mu^{\rm f_2^\prime-pole}=-\left(0.042(13)_{\rm stat}(^{+0.000}_{-0.008})_{\rm syst}\right)$, adding up to $a_\mu^{a_2+f_2+f_2^\prime \rm - pole}=-\left(4.3^{+0.3}_{-0.5}\right)$, agree very well with the holographic computation if $\mathcal{F}_3^T$ is set to zero, $-4.5(6)\times10^{-11}$~\cite{Cappiello:2025fyf}, and exceed in magnitude the outcome of the dispersive analysis, $-2.5(3)\times10^{-11}$~\cite{Hoferichter:2024bae}.

Then, we have extended our Lagrangian, including contributions generating $\mathcal{F}_3^T$, finding (
same units) $a_\mu^{\rm a_2-pole}=+0.47(1.43)_{\rm norm}(3)_{\rm stat}(^{+0.06}_{-0.00})_{\rm syst}$, $a_\mu^{\rm f_2-pole}=+1.18(4.18)_{\rm norm} (12)_{\rm stat}(^{+0.24}_{-0.00})_{\rm syst}$, $a_\mu^{\rm f_2'-pole}=+0.040(78)_{\rm norm}(2)_{\rm stat}$ , with uncertainty dominated by that on $\mathcal{F}_3^T(0,0)$. These add up to $a_\mu^{a_2+f_2+f_2^\prime \rm - pole}=+1.7(4.4)$, in agreement in sign with the analogous holographic result~\footnote{{Results in Refs. \cite{Cappiello:2025fyf,Mager:2025pvz}, involve additional contributions from excited tensor mesons, which would bring into agreement the phenomenological determination with the lattice result for the whole HLbL. The value that we quote concerns the pole contribution of the ground-state tensor mesons, as in this work.}},
and compatible within errors $+3.2(0.4)$~\cite{Cappiello:2025fyf}. 
{The positive sign, as in the holographic result, was expected as $\mathcal{F}_T^3$ has the same sign in both approaches, and compatible normalizations. However, in our approach, experimental information on $\mathcal{F}_3^T(0,0)$ would greatly improve our description of this form factor and its implications for $a_\mu$. This underlines the need to measure double-virtual data for the process $T\to\gamma^*\gamma^*$ as 
the helicity basis of eq. (\ref{eq:helicitybasis}) suggests.} We hope these results be helpful in a better understanding of the $a_\mu^{\rm T-poles}$ contribution and motivate the experimentalists to achieve these new form factor measurements. Finally, we remark that we have first pointed out how non-vanishing $\mathcal{F}_{2,4,5}^T$ appear. Computing their contributions to $a_\mu$ would require a new basis, which seems to be a future necessary development of the existing formalism, needed --in particular-- for a complete evaluation of $a_\mu^{\rm T-pole}$ within $R \chi T$.
\section*{Acknowledgements}
The authors are very grateful to Antonio Rodríguez-Sánchez, Anton Rebhan, and Jonas Mager for enlightening feedback and insightful comments. We also acknowledge extremely helpful correspondence with Luigi Cappiello and Martin Hoferichter. The authors thank partial funding from CONAHCYT and SECIHTI (México), particularly for E.~J.~E. Ph. D. scholarship. P.~R. acknowledges funding through the project CBF2023-2024-3226, as well as Spanish support during his sabbatical through projects MCIN/AEI/10.13039/501100011033, grants PID2020-114473GB- I00
and PID2023-146220NB-I00, and Generalitat Valenciana grant PROMETEO/2021/07.
\section*{Appendix}
\label{sec:appendix}
\begin{figure}
    \centering
    \includegraphics[width=1\linewidth]{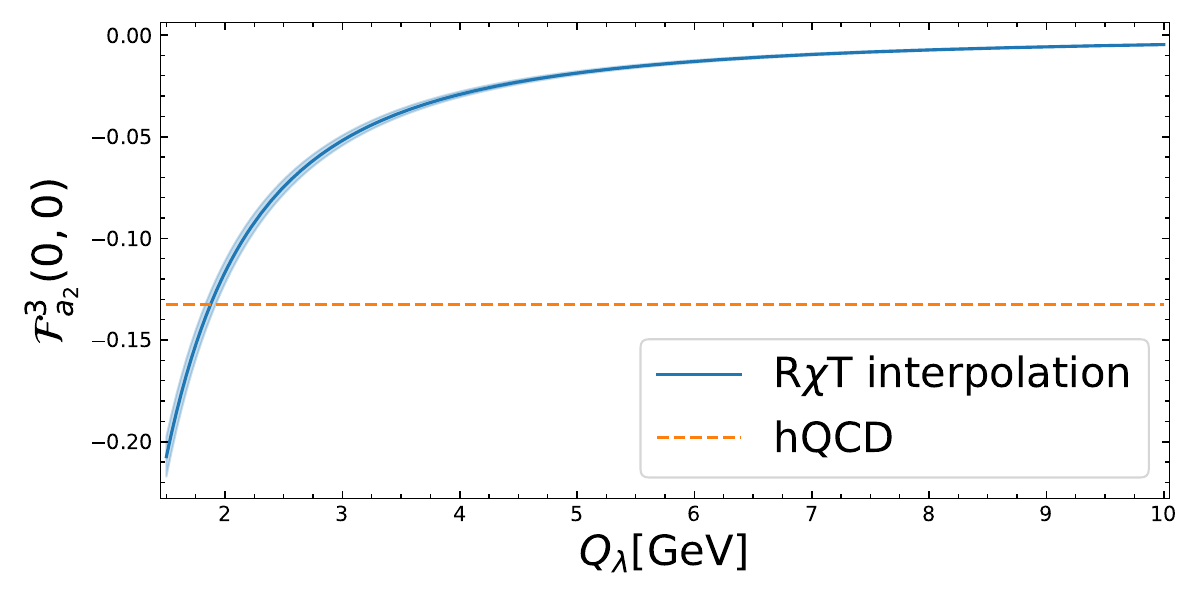}
    \caption{Inferred value of $\mathcal{F}_3^T(0,0)$ from $\mathcal{F}_1^T(0,0)$ within $R \chi T$ for all 3 neutral tensor mesons(blue) with our 1$\sigma$ error and the value given by the hQCD approach\cite{Cappiello:2025fyf}(orange).}
    \label{fig:F300infered}
\end{figure}

The lack of information on the normalization of the $\mathcal{F}_3^T$ form factor is the main problem for an $R \chi T$ description of them. In \cite{Cappiello:2025fyf}, a value for $\mathcal{F}_3^T(0,0)$ was obtained within holographic QCD. {We aim to estimate a reasonable ballpark of this value within $R \chi T$.} We performed an interpolation from the normalization of $\mathcal{F}_1^T(0,0)$. This procedure is based on the fulfillment of the SDCs coming from the Light-Cone-Expansion\cite{Hoferichter:2020lap} results at different scales. We require that the ratio of the $R \chi T$ and the LCE one is matched for $\mathcal{F}_1^T$ and $\mathcal{F}_3^T$ at a scale $Q_\lambda$. This matching depends on $\mathcal{F}_3^T(0,0)$, specifically for the $a_2$ case, since it defines the rest of the normalizations as can be seen from eqn. (\ref{eq:F3}). Consequently, floating the matching scale leads to different values for the normalization. We will take the average of the obtained upper and lower limits and use an error that spans the whole range of results for $\mathcal{F}^{a_2}_3(0,0)$. The matching condition is then:
\begin{equation}
    \frac{\mathcal{F}_3^{a_2-\rm R\chi T}(Q_\lambda^2,Q_\lambda^2)}{\mathcal{F}_3^{a_2-\rm LCE}(Q_\lambda^2,Q_\lambda^2)}=    \frac{\mathcal{F}_1^{a_2-\rm R\chi T}(Q_\lambda^2,Q_\lambda^2)}{\mathcal{F}_1^{a_2-\rm LCE}(Q_\lambda^2,Q_\lambda^2)}.
\end{equation}
The results for the matching for the 3 neutral tensor mesons are shown in Fig. \ref{fig:F300infered}. It can be seen that the value of the hQCD is included in the possible region for the normalization coming directly from the $R \chi T$ inference. This happens for all 3 tensor mesons. The allowed region is wide; however, we follow this conservative approach given the lack of information on the normalization of the FFs. This matching can be observed graphically by comparing Figures \ref{fig:matching1} and \ref{fig:matching3}.

As it can be seen, this procedure spans a wide range of possible \textit{negative} values for $\mathcal{F}_3^T(0,0)$. This sign implies that we do not expect a sign change in the form factor, as the asymptotic behavior is also negative and the function is monotonic{, as in the hQCD result}. The values for the normalizations are:
\begin{gather}
    \mathcal{F}_3^{a_2}(0,0)=-(0.106\pm0.101),\nonumber\\
    \mathcal{F}_3^{f_2}(0,0)=-(0.165\pm0.158),\nonumber\\
    \mathcal{F}_3^{f_2^\prime}(0,0)=-(0.038\pm0.036).\nonumber
\end{gather} 
\bibliographystyle{JHEP.bst}
\bibliography{Tpoles.bib}

\end{document}